\documentclass[preprintnumbers,amsmath,amssymb,twocolumn,showpacs]{revtex4}
\usepackage{graphicx}
\usepackage{dcolumn}
\usepackage{bm}

\usepackage{multirow}
\usepackage{graphicx}
\usepackage{hyperref}
\usepackage{amsmath}

\begin{document}

\title{Atom-atom entanglement by single-photon detection}

\author{L. Slodi\v{c}ka$^1$,  G. H\'etet$^{1}$, N. R\"ock$^1$, P. Schindler$^1$, M. Hennrich$^1$, and R. Blatt$^{1,2}$}

\date{\today}

\affiliation{
$^1$ Institut f{\"u}r Experimentalphysik, Universit{\"a}t Innsbruck, Technikerstra{\ss}e 25, 6020 Innsbruck, Austria
$^2$ Institut f\"ur Quantenoptik und Quanteninformation der \"Osterreichischen Akademie der Wissenschaften, Technikerstra{\ss}e 21a, 6020 Innsbruck, Austria}

\begin{abstract}
A scheme for entangling distant atoms is realized, as proposed in the seminal paper by Cabrillo et al. [Phys. Rev. A \textbf{59}, 1025 (1999)]. The protocol is based on quantum interference and detection of a single photon scattered from two effectively one meter distant laser-cooled and trapped atomic ions. The detection of a single photon heralds entanglement of two internal states of the trapped ions with high rate and with a fidelity limited mostly by atomic motion. Control of the entangled state phase is demonstrated by changing the path length of the single-photon interferometer.

\end{abstract}

\maketitle

The generation of entanglement between distant physical systems is
an essential primitive for quantum communication
networks \cite{Bri98,Dua01} and further tests of quantum mechanics.
The realization of heralded entanglement between distant atomic ensembles \cite{Cho05,Cho07}
was amongst the first major achievements in this direction.
Probabilistic generation of heralded entanglement between single
atoms \cite{Sim03} was demonstrated using single trapped ions \cite{Moe07} and 
neutral atoms \cite{Hof12} with an entanglement generation rate
given by the probability of coincident detection of the
two photons coming from the atoms \cite{Zip08,Luo09}.
More recently, single neutral atoms trapped at distant locations were entangled
by first generating the single atom-photon entanglement and then
mapping the photonic state on the electronic state of the second
atom \cite{Rit12}. A heralding mechanism will however be essential
for efficient entanglement and scalability of quantum networks using realistic
channels \cite{Dua01}, and single qubit operations are required for
distributed quantum information processing schemes \cite{Dua10}.
In this Letter we report on the realization of a fundamental
process which fulfills both these conditions by showing entanglement
between two well-defined atomic qubits via emission and detection of a single light quanta \cite{Cab99}. In this scheme, both the energy and the phase of the emitted single photon are used for entanglement generation. In addition, this mechanism allows the demonstration of a large speedup in entanglement generation rate compared to the previously
realized heralded entanglement protocol with single
atoms \cite{Zip08, Moe07}. This result will enable
the practical distribution of quantum information over long
distances using single atom architectures.

\begin{figure}
\begin{center}
\includegraphics[scale=0.075]{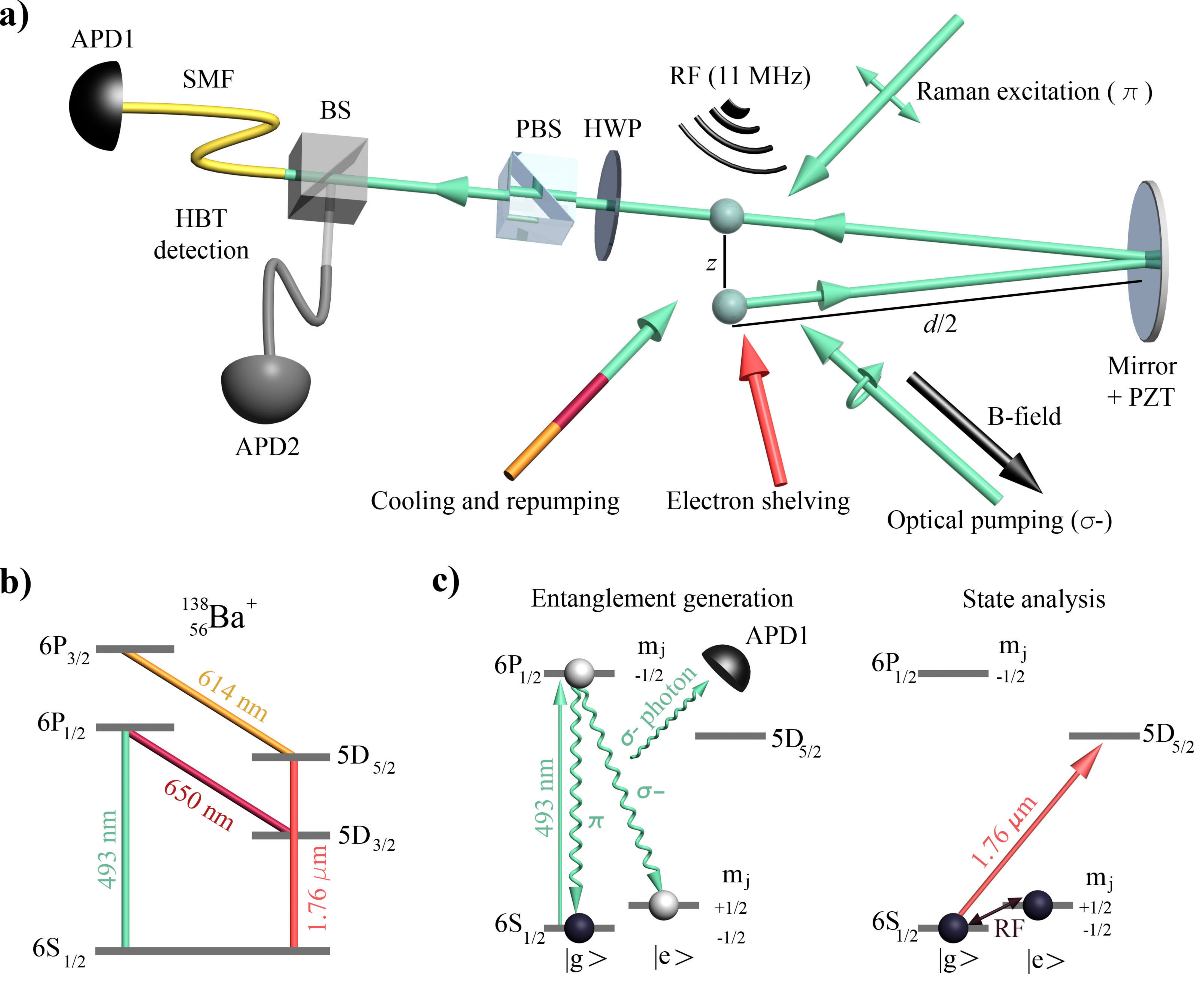}
\caption{Experimental procedure for entanglement
generation. a) The fluorescence of the two ions is overlapped
using a distant mirror which sets the effective distance between
them to $d=1$~meter. A half wave plate (HWP), a polarizing beam
splitter (PBS) and a single-mode optical fiber select the
polarization and the spatial mode before an avalanche photodiode
(APD1). A non-polarizing beam-splitter and an additional avalanche
photodiode (APD2) can be inserted to form a Hanbury-Brown-Twiss
setup. See details in main text. b) Level scheme of $^{138}$Ba$^+$
including the wavelengths of the lasers used in our experiment. c)
Experimental sequence. Spontaneous Raman scattering to $|e\rangle$
triggers emission of a single photon from the two atoms. Upon
successful detection of a $\sigma^-$ photon, state analysis
comprising coherent radio-frequency (RF) pulses at 11\,MHz, and
electron shelving to the 5D$_{5/2}$ level are performed. See
details in main text. \label{setup}}
\end{center}
\end{figure}

Entanglement of distant single atoms through the detection of a
single photon, as proposed in the seminal work of Cabrillo et al.
\cite{Cab99}, is both a fundamental and a promising technique for
the field of quantum information. The interconnection between
quantum nodes based on this scheme would provide efficient
distribution of quantum information in large scale quantum
networks \cite{Zip08,Luo09}. To generate heralded entanglement,
two atoms ($A$,$B$) are both prepared in the same long-lived
electronic state $|gg\rangle$. Each atom is excited with a small
probability $p_e$ to another metastable state $|e\rangle$ through
a spontaneous Raman process
($|g\rangle\rightarrow|i\rangle\rightarrow|e\rangle $) by weak
excitation of the $|g\rangle\rightarrow|i\rangle$ transition and
spontaneous emission of the single photon on the
$|i\rangle\rightarrow|e\rangle$. Here $|i\rangle$ denotes an
auxiliary atomic state with short lifetime. This Raman process
entangles each of the atom's internal states with the emitted
photon number, so the state of each atom and its corresponding
light mode can be written as
$\sqrt{1-p_e}|g,0\rangle e^{i \phi_{L}} +\sqrt{p_e}|e,1\rangle e^{i \phi_D}$. Here,
the phases $\phi_L$ and $\phi_D$ correspond to the phase of the exciting laser at the position of atom $A$ and the phase acquired by the spontaneously emitted photon on its way to the detector, respectively.
Indistinguishability of the photons from the two atoms is achieved
by overlapping their corresponding modes, for example using a beam
splitter. The total state of the system consisting of both atoms
and the light modes is then
$(1-p_e)e^{i(\phi_{L,A}+\phi_{L,B})}|gg,0\rangle+\sqrt{p_e(1-p_e)}(e^{i(\phi_{L,A}+\phi_{D,B})}|eg,1\rangle+e^{i(\phi_{L,B}+\phi_{D,A})}|ge,1\rangle)+p_e
e^{i (\phi_{D,A}+\phi_{D,B})}|ee,2\rangle$. Single photon detection projects the two-atom state onto an entangled state
$|\Psi^\phi\rangle=\frac{1}{\sqrt{2}}(|eg\rangle+e^{i\phi}|ge\rangle)$.
Since at least one atom must be excited, the probability of measuring such a state is then 1-$p_e^2$. Here $p_e^2$ is the probability of simultaneous excitation of both atoms. The absolute success probability of the entangled state generation in one experimental run is then $P_{\rm succ} = 2 p_e(1-p_e)\eta$, where $\eta$ is the overall detection efficiency of the generated photons. The phase of the generated entangled state $\phi$ corresponds to the sum of the phase difference acquired by
exciting beam at the position of the two atoms and the phase
difference acquired by the photons from the
respective atoms upon travelling to the detector. The only limiting factor here is the
probability of simultaneous excitation of the two atoms $p_e^2$,
which can be, in principle, made arbitrarily small.

To experimentally demonstrate the creation of such a single-photon
heralded entanglement two barium ions are trapped and cooled in a
linear Paul trap\cite{Slo12}. As shown in Fig.~\ref{setup}\,a) and b),
laser light at 493\,nm is used to Doppler-cool the ions and to
detect their electronic states by means of electron shelving,
and a laser field at 650\,nm repumps the atoms to the $P_{1/2}$
level from the metastable D$_{3/2}$ state. By carefully adjusting
the cooling and trapping parameters, the ions are always well
within the Lamb-Dicke limit so that the photon recoil during the
Raman scattering process is mostly carried by the trap. This ensures that only minimal
information is retained in the motion of the ion about which atom
scattered the photon during the entanglement generation process.
The fluorescence photons are efficiently collected by two high
numerical aperture lenses (NA $\approx 0.4$) placed 14\,mm away
from the atoms. A magnetic field of 0.4\,mT is applied at an angle
of 45\,degrees with respect to the two-ion axis and defines the
quantization axis. After passing through a polarizing beam
splitter that blocks the $\pi$-polarized light and lets
$\sigma$-polarized light pass, the spatial overlap of the photons
is guaranteed by collecting the atomic fluorescence of the first
ion in a single mode optical fiber, whilst the fluorescence of the
second ion is sent to a distant mirror that retro-reflects it in
the same optical fiber \cite{Ris08}. The fluorescence of the two
ions (including the Raman scattered light) is then detected by an
avalanche photodiode with a quantum efficiency of 60\%.

For efficient generation of the two-atom entangled state, the
emitted photons must be indistinguishable in all degrees of
freedom at the position of the triggering detector. We
characterize their indistinguishability by a measurement of the
first and second order correlation functions (see supplemental material A). These measurements yield unambiguous separation between the
major decoherence mechanisms and lead to the conclusion that
which-way information given by atomic motion is the main source of
distinguishability.

In the entanglement generation procedure, we first Doppler-cool
the ions and stabilize the mirror-ion distance $d/2$ by locking the position of the interference fringe measured during the
Doppler cooling sequence to a chosen position, see Fig.\ref{setup}-a). The ion internal
states are then prepared to the Zeeman substates $|6S_{1/2},
m_j=-1/2\rangle = |g\rangle$ by optical pumping with a circularly
polarized laser pulse propagating along the magnetic field. Then,
a weak horizontally polarized laser pulse (Raman excitation)
excites both ions on the S$_{1/2}\leftrightarrow$ P$_{1/2}$
transition with a probability $p_e=0.07$ through a resonant
spontaneous Raman scattering to the other Zeeman sublevel
(m$_j$=+1/2) of the 6S$_{1/2}$ state, $|e\rangle$.
The electronic state of each ion is at this point entangled with
the number of photons $|0\rangle$ or $|1\rangle$ in the $\sigma^-$
polarized photonic mode. Provided that high indistinguishability
of the two photonic channels is assured and that simultaneous
excitation of both atoms is negligible, detection of a single
$\sigma^-$ photon on the APD projects the two-ion state
onto the maximally entangled state
$|\Psi\rangle=\frac{1}{\sqrt{2}}(|ge\rangle +|eg\rangle e^{i k
d})$, where $k$ is the wavenumber of the 493\,nm fluorescence. The phase factor $e^{i k
d}$ corresponds here solely to the phase difference $\phi_{D,A}-\phi_{D,B}$ acquired by the emitted photon upon its way to the detector. The phase difference of the excitation laser at the position of two ions $\phi_{L,A}-\phi_{L,B}$ is fixed to $n\times 2\pi, n\in I$ by setting the mutual distance between the ions in the trap to $z=n \lambda/\cos\theta$, where $\theta$ is the angle between the Raman-excitation laser direction and the ion-crystal axis. We will first demonstrate a successful preparation of the Bell state
$|\Psi^+\rangle$ for the phase $e^{i k d}=1$ corresponding to an
antinode of the interference fringe.

Following the detection of a Raman scattered $\sigma^-$ photon, we
coherently manipulate the generated two-atom state to allow for
measurements in a different bases. As shown in Fig.\ref{setup}-c),
this is done by first applying radio-frequency (RF) pulses that
are resonant with the $|g\rangle \leftrightarrow |e\rangle$
transition of both atoms. Discrimination between the two Zeeman
sub-levels of the S$_{1/2}$ state is finally done by shelving the
population of the m$_j=-1/2$ state to the metastable D$_{5/2}$
level using a narrowband 1.76\,$\mu$m laser \cite{Slo12}. The fluorescence on the S$_{1/2}\leftrightarrow$P$_{1/2}$ transition allows us to measure the two-atom state. By setting the appropriate thresholds on the fluorescence counting histogram, we can discriminate the three possible cases where no excitations are present in the two atoms, a single excitation is shared between the two atoms, and where two atoms are excited. These events can all be separated with 98\% probability,
enabling us to efficiently reconstruct the relevant parts of the
density matrix of the two-atom state. The 614\,nm laser field then
resets the ions to the 6S$_{1/2}$ state and the same experiment is
repeated 100 times.

\begin{figure}
\begin{center}
\includegraphics[scale=1.9]{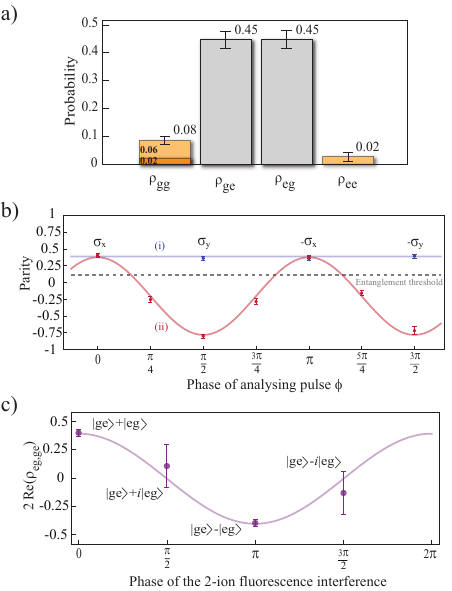}
\caption{Characterization of the entangled state. a)
Two-atom state populations after the detection of a $\sigma^-$
photon showing that the total probability of measuring the state
with a single excitation is 90\%. b) Parity measurements as a
function of the RF-phase. Trace (ii) corresponds to the
measurement of the atomic populations after two global rotations
$\hat{R}^g(\pi/2,\phi)\hat{R}^g(\pi/2,\pi/2)$. In the measurement
of trace (i) only a single global RF-pulse $\hat{R}^g(\pi/2,\phi)$
is applied. The dashed line shows the threshold for entanglement,
estimated from the measured diagonal terms. c) Real
part of the coherence between the $|ge\rangle$ and $|eg\rangle$
states as a function of the phase of the optical path difference
between the two ions.\label{Parity}}
\end{center}
\end{figure}

Fig.~\ref{Parity}-a) shows the measurement results obtained
without the RF analysis pulses. The results tell us that 89$\pm
3$\% of all the triggering events signal that only one of the
atoms was excited to the $|e\rangle$ state. The remaining 10\%
errors are caused by APD dark counts and double excitation of the
ions. Our detection process using a single photomultiplier doesn't
allow us to resolve individual $\rho_{eg}$ and $\rho_{ge}$
populations directly, but it tells us the number of the excited
atoms, so the sum of these terms. Although individual populations
of the $\rho_{eg}$ and $\rho_{ge}$ states are not needed for
estimation of the fidelity with the state $|\Psi^+\rangle$, we
also experimentally prove that $\rho_{eg}$ and $\rho_{ge}$
populations are approximately the same and depend only on the
overall fluorescence detection efficiencies from the two ions. In
order to measure the quantum coherence of the generated state, we
then apply two consecutive global RF-pulses, each corresponding to
the rotation
$\hat{R}(\theta,\phi)=\exp{(-i\frac{\theta}{2}(\cos\phi\hat{S}_x+\sin\phi\hat{S}_y))}$,
where
$\hat{S}_{x,y}=\hat{\sigma}_{x,y}^{(1)}\otimes\hat{\sigma}_{x,y}^{(2)}$
is the global Pauli operator
acting on both ions. The rotation angle $\theta$ and rotation axis
$\phi$ on the Bloch sphere are determined by the duration and the
phase of the RF pulses, respectively. We first apply the pulse
$\hat{R}(\pi/2,\pi/2)$ which performs the unitary rotation
$\hat{R}(\pi/2,\pi/2)|\Psi^+\rangle \rightarrow |\Phi^-\rangle$,
where $|\Phi^-\rangle=\frac{1}{\sqrt{2}}(|gg\rangle-|ee\rangle)$.
A second RF-pulse with same duration but with a phase $\phi$ then
performs the rotation $\hat{R}(\pi/2,\phi)|\Phi^-\rangle$. After
shelving the state $|e\rangle$ to the metastable level $D_{5/2}$,
we scatter light from both ions on the cooling transition. From
the measured fluorescence rate at different phases $\phi$, we
extract the mean value of the parity operator defined as
$\hat{P}=\hat{p}_{gg} +\hat{p}_{ee}-\hat{p}_{eg}-\hat{p}_{ge}$,
where $\hat{p}_{ij}$ are the projection operators on states
$|ij\rangle$, $i,j\in \{g,e\}$ \cite{Sac00} (see supplemental materials B).

Fig.~\ref{Parity}-b), trace (ii), shows the results of the parity
operator measurements preceded by two global RF rotations
$\hat{R}(\pi/2,\phi)\hat{R}(\pi/2,\pi/2)$. The measured parity
clearly oscillates as a function of phase $\phi$ with contrast of
58.0$\pm 2.5$\% and period of $\pi$, a proof that we indeed succeed in preparing an
entangled two-ion state close to $|\Psi^+\rangle$ \cite{Sac00}.
The mean value of the parity operator at zero phase
$\langle\hat{P}\rangle_{\phi\rightarrow 0}$ corresponds to the
difference between the inner parts and outer-most coherence terms
of the density matrix. We evaluate it to be
$2\rm{Re}(\rho_{ge,eg}-\rho_{gg,ee})=0.38\pm 0.03$. To precisely
quantify the fidelity of our state with $|\Psi^+\rangle$, we
however need to estimate the real part of the coherence
$\rho_{ge,eg}$ itself. This is done by measuring the parity
without the first RF rotation. Trace (i) of Fig.~\ref{Parity}-b)
shows the expectation value of the parity as a function of the
phase $\phi$ of the single RF-pulse $\hat{R}(\pi/2,\phi)$. The only oscillatory term contributing to this parity measurement reads $2 (\sin(2\phi)\rm{Im}\rho_{gg,ee}-\cos(2\phi)\rm{Re}\rho_{gg,ee})$. The measured data however shows independence of the parity signal with respect to the phase $\phi$ within the measurement error. Therefore, only the coherence corresponding to the state $|\Psi^+\rangle$ contributes to the parity signal (ii). The value of the coherence $\rm{Re}(\rho_{gg,ee})$ estimated from these measurements is $\rm{Re}(\rho_{gg,ee})=0.00\pm 0.03$.
We finally estimate that the fidelity of the generated state with the maximally entangled state $|\Psi^+\rangle$ is $F=64\pm 2\%$. The threshold for an entanglement is thus surpassed by more than six standard deviations.

The coherence between the $|ge\rangle$ and $|eg\rangle$ states of
$38\pm 3\%$ is limited by three main processes. First, imperfect
populations of $|ge\rangle$ and $|eg\rangle$ states set a limit of
89\% \cite{Shi06}. Around 4\% of the coherence loss can be
attributed to the finite coherence time of the individual atomic
qubits (120$\mu$s) due to collective magnetic field fluctuations.
Although the generated $|\Psi^+\rangle$ state is intrinsically
insensitive against collective dephasing \cite{Kie01,Roo04}, a
loss of coherence is indeed expected after the rotation of
$|\Psi^+\rangle$ out of the decoherence-free subspace. The highest
contribution to the coherence loss can be attributed to atomic
motion, which can provide information about which atom emitted the
photon. Around 55\% of the coherence is lost due to the atomic recoil
kicks during the Raman scattering (see supplemental material C). Errorbars in the presented measurements
results correspond to one standard deviation and are estimated
statistically from several experimental runs each giving
approximately 120 measurement outcomes. Up to 60\,\% of the
measurement error is caused by the quantum projection noise.
Additional uncertainty comes from slow magnetic field drifts with
a magnitude of several tens of nT making the RF-driving
off-resonant by tens of kHz.

An intrinsic feature of the realized entangling protocol is the
dependence of the generated entangled state phase on the optical
path difference between the ions. To demonstrate this, we measure
the real part of the coherence between the $|ge\rangle$ and
$|eg\rangle$ states as a function of the phase factor $k d$.
Fig.~\ref{Parity}-c) reveals a large change of the real part of
the coherence from positive to negative values when going from the
maximum to the minimum of the interference signal, in agreement
with the $e^{i k d}$ phase dependence of the entangled state.

An important feature of the single-photon heralding mechanism is
the high entanglement generation rate that can be achieved. With
our experimental set-up, the single photon detection
scheme indeed yields a higher rate compared to the two photon
scheme proposed by Simon et al. \cite{Sim03,Zip08,Hof12,Mat08}. The probability
of preparing an entangled state depends on the probability of the
single photon detection and the Raman scattering
probabilities \cite{Zip08}, which in our case gives a total of
$P_{\rm succ}=1.1\times 10^{-4}$ for each trial run. With an experimental duty cycle of
2.3\,kHz, this corresponds to 15.4 successful entanglement
generation events/minute, which is in good agreement with the
experimentally observed 14 $\pm 2$ events/minute. A detailed
analysis of the overall photon-detection efficiency can be found
in the supplemental material E.

We have demonstrated a fundamentally new protocol for generating heralded entanglement between two ions.
This was achieved via the scheme proposed in the seminal work of
Cabrillo et al. \cite{Cab99} where two atoms are entangled with the emission and detection of only one photon. Such a single-photon scheme allowed us to reach a rate of entanglement generation of 14\,events/minute, more than two orders of magnitude higher
than the rate obtainable with protocols relying on a two-photon coincidence
events with our experimental parameters. The maximally entangled state
$|\Psi^+\rangle$ is produced with a fidelity of 63.5\,\% limited
mostly by residual atomic motion. These results can be
improved by cooling all of the involved motional modes close to
their ground state \cite{Slo12} or choosing a different excitation
direction to minimize residual which-way information. These improvements, together with the experimental results presented, will enable efficient creation and distribution of entanglement between distant sites with well-defined and controllable atomic qubits. Such entanglement generation corresponds to an essential building block of scalable quantum communication \cite{Dua10} and distributed quantum computation \cite{Jia07,Cir99,Got99} architectures with single atoms.

We thank J. Eschner, T. Monz, T. Northup and A. Stute
for helpful discussions. This work was supported by the Austrian
Science Fund (FWF), the Institut f\"ur Quanteninformation GmbH,
and a Marie Curie International Incoming Fellowship within the 8th
European Framework Program.


\begin{thebibliography}{10}
\expandafter\ifx\csname url\endcsname\relax
  \def\url#1{\texttt{#1}}\fi
\expandafter\ifx\csname urlprefix\endcsname\relax\def\urlprefix{URL }\fi
\providecommand{\bibinfo}[2]{#2}
\providecommand{\eprint}[2][]{\url{#2}}



\bibitem{Bri98}
\bibinfo{author}{Briegel, H.~J.} \emph{et~al.}
\newblock \emph{\bibinfo{journal}{Phys. Rev. Lett.}} \textbf{\bibinfo{volume}{81}},
  \bibinfo{pages}{5932–5935} (\bibinfo{year}{1998}).

\bibitem{Dua01}
\bibinfo{author}{Duan, L.~M.} \emph{et~al.}
\newblock \emph{\bibinfo{journal}{Nature}} \textbf{\bibinfo{volume}{414}},
  \bibinfo{pages}{413-418} (\bibinfo{year}{2001}).

\bibitem{Cho05}
\bibinfo{author}{Chou, C. W.} \emph{et~al.}
\newblock \emph{\bibinfo{journal}{Nature}}
  \textbf{\bibinfo{volume}{438}}, \bibinfo{pages}{828}
  (\bibinfo{year}{2005}).

\bibitem{Cho07}
\bibinfo{author}{Chou, C. W.} \emph{et~al.}
\newblock \emph{\bibinfo{journal}{Science}}
  \textbf{\bibinfo{volume}{316}}, \bibinfo{pages}{1316}
  (\bibinfo{year}{2007}).


\bibitem{Sim03}
\bibinfo{author}{Simon, C.} \& \bibinfo{author}{Irvine, W. T.~M.}
\newblock \emph{\bibinfo{journal}{Phys. Rev. Lett.}}
  \textbf{\bibinfo{volume}{91}}, \bibinfo{pages}{110405}
  (\bibinfo{year}{2003}).



\bibitem{Moe07}
\bibinfo{author}{Moehring, D.~L.} \emph{et~al.}
\newblock \emph{\bibinfo{journal}{Nature}} \textbf{\bibinfo{volume}{449}},
  \bibinfo{pages}{68--71} (\bibinfo{year}{2007}).

\bibitem{Hof12}
\bibinfo{author}{Hofmann, J.} \emph{et~al.}
\newblock \emph{\bibinfo{journal}{Science}} \textbf{\bibinfo{volume}{337}},
  \bibinfo{pages}{72--75} (\bibinfo{year}{2012}).


\bibitem{Zip08}
\bibinfo{author}{Zippilli, S.} \emph{et~al.}
\newblock \emph{\bibinfo{journal}{New Journal of Physics}}
  \textbf{\bibinfo{volume}{10}}, \bibinfo{pages}{103003}
  (\bibinfo{year}{2008}).


\bibitem{Luo09}
\bibinfo{author}{Luo, L.} \emph{et~al.}
\newblock \emph{\bibinfo{journal}{Fortschritte der Physik}}
  \textbf{\bibinfo{volume}{57}}, \bibinfo{pages}{1133--1152}
  (\bibinfo{year}{2009}).


\bibitem{Rit12}
\bibinfo{author}{Ritter, S.} \emph{et~al.}
\newblock \emph{\bibinfo{journal}{Nature}}
  \textbf{\bibinfo{volume}{484}}, \bibinfo{pages}{195–200}
  (\bibinfo{year}{2012}).

\bibitem{Dua10}
\bibinfo{author}{Duan, L. -M.} \& \bibinfo{author}{Monroe, C.}
\newblock \emph{\bibinfo{journal}{Rev. Mod. Phys.}}
  \textbf{\bibinfo{volume}{82}}, \bibinfo{pages}{1209--1224}
  (\bibinfo{year}{2010}).

\bibitem{Cab99}
\bibinfo{author}{Cabrillo, C.}, \bibinfo{author}{Cirac, J.~I.},
  \bibinfo{author}{Garcia-Fern\'andez, P.}  \& \bibinfo{author}{Zoller, P.}
\newblock \emph{\bibinfo{journal}{Phys. Rev. A}} \textbf{\bibinfo{volume}{59}},
  \bibinfo{pages}{1025--1033} (\bibinfo{year}{1999}).

\bibitem{Slo12}
\bibinfo{author}{Slodi\ifmmode~\check{c}\else \v{c}\fi{}ka, L.} \emph{et~al.}
\newblock \emph{\bibinfo{journal}{Phys. Rev. A}} \textbf{\bibinfo{volume}{85}},
  \bibinfo{pages}{043401} (\bibinfo{year}{2012}).


\bibitem{Ris08}
\bibinfo{author}{We note that in our setup,
super/sub-radiance effects that might occur here are on the order
of $1\%$, and can thus be neglected. See Rist, S.} \emph{et~al.}
\newblock \emph{\bibinfo{journal}{Phys. Rev. A}} \textbf{\bibinfo{volume}{78}},
  \bibinfo{pages}{013808} (\bibinfo{year}{2008}).

%
%

\bibitem{Sac00}
\bibinfo{author}{Sackett, C.~A.} \emph{et~al.}
\newblock \emph{\bibinfo{journal}{Nature}} \textbf{\bibinfo{volume}{404}},
  \bibinfo{pages}{256--258} (\bibinfo{year}{2000}).

\bibitem{Shi06}
\bibinfo{author}{Shirokov, M.} \emph{et~al.}
\newblock \emph{\bibinfo{journal}{International Journal of Theoretical Physics}} \textbf{\bibinfo{volume}{45}},
  \bibinfo{pages}{141--151} (\bibinfo{year}{2006}).

\bibitem{Roo04}
\bibinfo{author}{Roos, C.~F.} \emph{et~al.}
\newblock \emph{\bibinfo{journal}{Phys. Rev. Lett.}} \textbf{\bibinfo{volume}{92}},
  \bibinfo{pages}{220402} (\bibinfo{year}{2004}).



\bibitem{Kie01}
\bibinfo{author}{Kielpinski, D.} \emph{et~al.}
\newblock \emph{\bibinfo{journal}{Science}} \textbf{\bibinfo{volume}{291}},
  \bibinfo{pages}{1013--1015} (\bibinfo{year}{2001}).


\bibitem{Mat08}
\bibinfo{author}{Matsukevich, D. N.} \emph{et~al.}
\newblock \emph{\bibinfo{journal}{Phys. Rev. Lett.}} \textbf{\bibinfo{volume}{100}},
  \bibinfo{pages}{150404} (\bibinfo{year}{2008}).  


\bibitem{Jia07}
\bibinfo{author}{Jiang, L.} \emph{et~al.}
\newblock \emph{\bibinfo{journal}{Phys. Rev. A}} \textbf{\bibinfo{volume}{76}},
  \bibinfo{pages}{062323} (\bibinfo{year}{2007}).


\bibitem{Cir99}
\bibinfo{author}{Cirac, J. I.} \emph{et~al.}
\newblock \emph{\bibinfo{journal}{Phys. Rev. A}} \textbf{\bibinfo{volume}{59}},
  \bibinfo{pages}{4249–-4254} (\bibinfo{year}{1999}).


\bibitem{Got99}
\bibinfo{author}{Gottesman, D.} \& \bibinfo{author}{Chuang, I. L.}
\newblock \emph{\bibinfo{journal}{Nature}} \textbf{\bibinfo{volume}{402}},
  \bibinfo{pages}{390--393} (\bibinfo{year}{1999}).


\end{thebibliography}

\begin{thebibliography}{10}
\expandafter\ifx\csname url\endcsname\relax
  \def\url#1{\texttt{#1}}\fi
\expandafter\ifx\csname urlprefix\endcsname\relax\def\urlprefix{URL }\fi
\providecommand{\bibinfo}[2]{#2}
\providecommand{\eprint}[2][]{\url{#2}}




\bibitem{Eic93}
\bibinfo{author}{Eichmann, U.} \emph{et~al.}
\newblock \emph{\bibinfo{journal}{Phys. Rev. Lett.}} \textbf{\bibinfo{volume}{70}},
  \bibinfo{pages}{220402} (\bibinfo{year}{2004}).

\bibitem{Ita98}
\bibinfo{author}{Itano, W. M.} \emph{et~al.}
\newblock \emph{\bibinfo{journal}{Phys. Rev. A}} \textbf{\bibinfo{volume}{57}},
  \bibinfo{pages}{4176--4187} (\bibinfo{year}{1998}).


\bibitem{Esc01}
\bibinfo{author}{Eschner, J.}, \bibinfo{author}{Raab, C.},
  \bibinfo{author}{Schmidt-Kaler, F.} \& \bibinfo{author}{Blatt, R.}
\newblock \emph{\bibinfo{journal}{Nature}} \textbf{\bibinfo{volume}{413}},
  \bibinfo{pages}{495--498} (\bibinfo{year}{2001}).

\bibitem{Slo12}
\bibinfo{author}{Slodi\ifmmode~\check{c}\else \v{c}\fi{}ka, L.} \emph{et~al.}
\newblock \emph{\bibinfo{journal}{Phys. Rev. A}} \textbf{\bibinfo{volume}{85}},
  \bibinfo{pages}{043401} (\bibinfo{year}{2012}).

\bibitem{Sac00}
\bibinfo{author}{Sackett, C.~A.} \emph{et~al.}
\newblock \emph{\bibinfo{journal}{Nature}} \textbf{\bibinfo{volume}{404}},
  \bibinfo{pages}{256--258} (\bibinfo{year}{2000}).

\bibitem{Cab99}
\bibinfo{author}{Cabrillo, C.}, \bibinfo{author}{Cirac, J.~I.},
  \bibinfo{author}{Garcia-Fern\'andez, P.} \& \bibinfo{author}{Zoller, P.}
\newblock \emph{\bibinfo{journal}{Phys. Rev. A}} \textbf{\bibinfo{volume}{59}},
  \bibinfo{pages}{1025--1033} (\bibinfo{year}{1999}).


\bibitem{Mon11}
\bibinfo{author}{Monz, T.} \emph{et~al.}
\newblock \emph{\bibinfo{journal}{Phys. Rev. Lett.}} \textbf{\bibinfo{volume}{106}},
  \bibinfo{pages}{130506} (\bibinfo{year}{2011}).



\bibitem{Sim03}
\bibinfo{author}{Simon, C.} \& \bibinfo{author}{Irvine, W. T.~M.}
\newblock \emph{\bibinfo{journal}{Phys. Rev. Lett.}}
  \textbf{\bibinfo{volume}{91}}, \bibinfo{pages}{110405}
  (\bibinfo{year}{2003}).
  
  


\end{thebibliography}

\newpage

\section*{\large{Supplemental material}}

\subsection{Indistinguishability measurements}

\begin{figure}
\begin{center}
\includegraphics[scale=1.4]{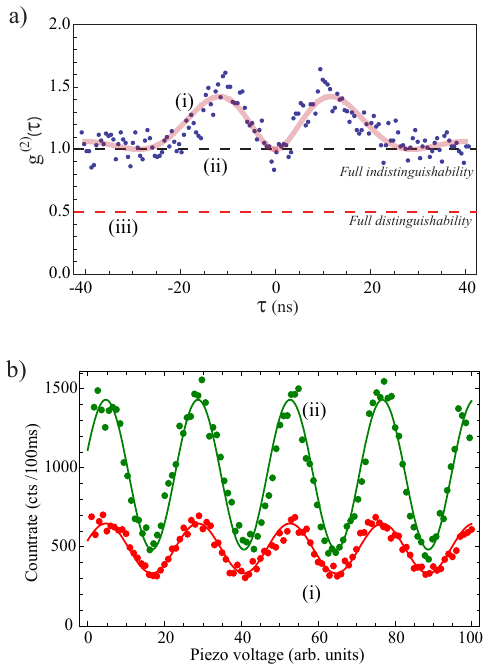}
\caption{Correlation functions measurements. a) Second
order correlation function of the two ions.  Trace (i) shows the
experimental results and theoretical fit. Trace (ii) and (iii) are
the expected values at $\tau=0$ for fully
indistinguishable/distinguishable emitters, respectively. b)
Interference fringes when the fluorescence of one ion overlaps
with itself (trace (ii)) and when the fluorescence light of both
ions is superimposed (trace (i)), in the limit of a weak
excitation, as a function of the phase difference between the two
optical paths. The contrasts are 60\% and 33\% respectively, both
limited mostly by atomic motion. \label{Temp_Spat_Coherence}}
\end{center}
\end{figure}

We estimate the degree of indistinguishability of photons emitted by the two atoms by a measurement of the
first and second order correlation functions.
Fig.~\ref{Temp_Spat_Coherence}-a), trace (i) shows the
second-order correlation function of the scattered light measured
using the Hanbury-Brown-Twiss detection setup.
At zero time delay between two consecutive
clicks, $g^{(2)}(\tau=0)=0.98\pm 0.07$ close to the theoretical value
1, a signature of high indistinguishability of the spatial and
polarization degrees of freedom in this optical mode (see supplemental material D). In order to estimate the amount of
motion induced which-way information, we measure the first order
correlation function. Fig.~\ref{Temp_Spat_Coherence}-b), trace (i)
shows the fluorescence intensity as a function of the distance $d$
between the two ions in a regime where elastic scattering
dominates. For comparison, trace (ii) shows the interference of
the fluorescence of one ion with itself under the same cooling
conditions. A contrast of up to 40\% is observed for the
interference of the elastic light scattered by the two ions.
Temporal decoherence caused by the photon emission and absorption
recoils \cite{Slo12,Eic93,Ita98,Esc01} is high enough to explain
this contrast. The measurement of the first and second order
correlation functions yields unambiguous separation between the
major decoherence mechanisms and leads to the conclusion that
which-way information given by atomic motion is the main source of
distinguishability.

\subsection{Parity measurements}

To characterize the maximally entangled state $|\Psi^+\rangle$, we perform a
Ramsey interference process that probes the coherence between the
$|ge\rangle$ and $|eg\rangle$ states.
In our experiment this is realized by a measurement of the parity
operator $\hat{P}$ in different bases.

The general state of our two qubit system can be described by the
$4\times 4$ hermitian matrix
\begin{equation}
\hat{\rho}=\begin{pmatrix}
\rho_{gg} & \rho_{gg,eg} & \rho_{gg,ge} & \rho_{gg,ee}\\
\rho_{gg,eg}^\ast & \rho_{eg} & \rho_{eg,ge} & \rho_{eg,ee}\\
\rho_{gg,ge}^\ast & \rho_{eg,ge}^\ast & \rho_{ge} & \rho_{ge,ee}\\
\rho_{gg,ee}^\ast & \rho_{eg,ee}^\ast & \rho_{ge,ee}^\ast & \rho_{ee}\\
\end{pmatrix}
\end{equation}
Measurement of the parity operator $\hat{P}$ on state $\hat{\rho}$
preceded by the collective rotation $\hat{R}(\theta,\phi)$ can
then be formally written as
Tr$[\hat{P}\hat{R}(\theta,\phi)\hat{\rho}(\hat{R}(\theta,\phi))^+]$.

For the $|\Psi^+\rangle$ state, it can be
readily shown that
\begin{equation}
{\rm
Tr}(\hat{P}\hat{R}(\pi/2,\phi)|\Psi^+\rangle\langle\Psi^+|(\hat{R}(\pi/2,\phi))^+)=1
\end{equation}
for all $\phi$. Parity measurement on the $|\Psi^+\rangle$
entangled state is therefore invariant with respect to the change
of the rotation pulse $\hat{R}(\pi/2,\phi)$ phase $\phi$. In order
to measure the parity oscillations for this state, we first have
to rotate by a global $\hat{R}(\pi/2,\pi/2)$ pulse, corresponding
to a $\hat{\sigma}_y$ rotation on both qubits with the pulse area
of $\pi/2$.

It can be shown that a peak-to-peak value of the parity
measurement oscillation higher than one with an oscillation period of $\phi=\pi$ on two qubits
is a sufficient condition for proving that the measured bi-partite
system is entangled \cite{Sac00}. To quantify the amount of
entanglement, we evaluate the fidelity $F=\langle
\Psi^+|\rho|\Psi^+\rangle$ with the maximally entangled state
$|\Psi^+\rangle$. It reads
\begin{equation}
F=\frac{1}{2} [\rho_{ge}+\rho_{eg}+ 2 {\rm Re}(\rho_{eg,ge})].
\label{fidelityRho}
\end{equation}
The fidelity thus depends only on the diagonal populations
$\rho_{ge}$ and $\rho_{eg}$ and on the real part of the
off-diagonal term $\rho_{eg,ge}$ that expresses the mutual
coherence between them. All these terms can be accessed using the
collective rotations $\hat{R}$ followed by the parity operator
measurement. Diagonal terms $\rho_{ge}$ and $\rho_{eg}$ can be
estimated directly by measuring the populations without any prior
RF-pulse application. The coherence term $\rho_{eg,ge}$ however
cannot be measured with a single global pulse sequence, because it
is always measured together with the coherence term
$\rho_{gg,ee}$. A simple way to separate their respective
contributions to the measured parity signal is to measure the
parity operator value after a single rotation
$\hat{R}(\pi/2,\phi)$ for different phases $\phi$. Independence of
the measured parity value on the phase $\phi$ proves
that the only coherence term contributing to the measured
coherence signal is the desired $\rho_{eg,ge}$. Table
\ref{tablePulses} shows examples of some relevant RF-pulses
rotation sequences and the corresponding parts of the density
matrix $\rho$ contributing to the measured signal.

\begin{table}
\begin{center}
\begin{tabular}{|c|c|c|}
\hline Number & Pulse & Measurement\\
of pulses & sequence & result\\
\hline \hline  0 & -- & $\rho_{gg}+\rho_{ee}-(\rho_{eg}+\rho_{ge})$\\
\hline \multirow{3}{*}{1} & $\hat{R}(\pi/2,0)$ & $2 \rm{Re}(\rho_{ge,eg}-\rho_{gg,ee})$\\
  & $\hat{R}(\pi/2,\pi/4)$ & $2 (\rm{Re}(\rho_{ge,eg})+\rm{Im}(\rho_{gg,ee}))$\\
  & $\hat{R}(\pi/2,\pi/2)$ & $2 \rm{Re}(\rho_{ge,eg}+\rho_{gg,ee})$\\
\hline \multirow{2}{*}{2} & $\hat{R}(\pi/2,0)\hat{R}(\pi/2,\pi/2)$ & $2 \rm{Re}(\rho_{ge,eg}-\rho_{gg,ee})$\\
  & $\hat{R}(\pi/2,\pi/2)\hat{R}(\pi/2,\pi/2)$&$\rho_{gg}+\rho_{ee}-(\rho_{eg}+\rho_{ge})$\\
\hline
\end{tabular}
\caption{Examples of the measurement sequences.
Expectation value of the parity operator after applying various
global RF-pulse sequences to the ions.} \label{tablePulses}
\end{center}
\end{table}

\subsection{Quantum coherence of the generated state}

The fidelity of the maximally entangled state $|\Psi^+\rangle$ with
the experimentally generated one, is given by \cite{Cab99}
\begin{eqnarray}\label{fidelity}
F&=&\frac{1}{2}\kappa(1 + F_{\rm dyn} e^{-4 t/\tau}).
\end{eqnarray}
Here $\kappa$ is a factor taking into account imperfect
populations of the $|ge\rangle$ and $|eg\rangle$ states mimicked
mostly by the detector dark-counts and double-excitations of ions
caused by imperfect setting of the Raman-beam polarization and
finite value of the excitation probability $p_e$. The latter gives
a double-excitation rate of $3 p_e^2\approx 1.5 \times 10^{-2}$.
$F_{\rm dyn}$ describes the decoherence due to atomic motion
(dynamical fidelity factor) and $ e^{-4 t/\tau}$ expresses the
loss of coherence due to the finite coherence time $\tau$ of each
individual qubit \cite{Mon11}. Here, $t$ corresponds to the time
relevant for this decoherence process, in our case this is the
time after the first pulse rotates the generated state out of the
decoherence free subspace to the analyzing pulse. The overall
fidelity $F$ is related to the respective density matrix elements
as defined in~(\ref{fidelity}), through
\begin{center}
\begin{eqnarray}
\rho_{ge}+\rho_{eg}= \kappa,\\
2 {\rm Re}(\rho_{eg,ge})=  \kappa F_{\rm dyn}e^{-4 t/\tau}.
\end{eqnarray}
\end{center}

The main factor contributing to decoherence is atomic motion.
One can show that
\begin{eqnarray}\label{Fdyn}
F_{\rm dyn}&=& e^{\langle [\vec{q}_1\cdot \vec{u}_1 - \vec{q}_2
\cdot \vec{u}_2]^2\rangle/2},
\end{eqnarray}
where $\vec{q}_{1,2}=\vec{k}_{\rm out}^{1,2}-\vec{k}_{\rm in}$ and
$\vec{k}_{\rm out}^{1} \approx -\vec{k}_{\rm out}^{2}$.
$\vec{k}_{\rm out,in}^{1,2}$ is the wavevector of the 493\,nm
light driving (in) and emitted (out) by atom 1 and 2
respectively. $\vec{u}_{1,2}=\vec{R}_{1,2}-\vec{R}^O_{1,2}$ are the displacements of atom 1 and 2 away from their equilibrium positions $\vec{R}^O_{1,2}$.
Decomposing the ion crystal motion into the normal
modes, one gets
\begin{equation}\label{Fdyn2}
\vec{q}_1\cdot \vec{u_1}-\vec{q_2} \cdot \vec{u_2}=-2k_{\rm
out}\hat{r}_{\rm cm}^{\rm rad}+2k_{\rm in}\cos{\phi}\hat{r}_{\rm
rel}^{\rm rad}-2k_{\rm in}\sin{\phi}\hat{r}_{\rm rel}^{\rm ax}
\end{equation}
where $\hat{r}_{\rm cm,rel}^{\rm rad,ax}$ are the position
operators of the quantized harmonic oscillator modes of the
two-ion crystal. ${\rm cm}$ and ${\rm rel}$ denote the center of
mass and stretch/rocking modes, respectively, and ${\rm rad, ax}$
are the radial and axial coupled modes. $\phi$ is the angle (in
our case $\pi/4$) between the Raman excitation laser and the
two-ion crystal axis. Inserting the expression\,\ref{Fdyn2} into
Eq.\,\ref{Fdyn}, we then get
\begin{eqnarray}
F_{\rm dyn}=e^{-2(k\sigma)^2},
\end{eqnarray}
where
\begin{eqnarray}
\sigma=\sqrt{(\sigma_{\rm cm}^{\rm rad})^2+\frac{1}{2}(\sigma_{\rm
rel}^{\rm rad})^2+\frac{1}{2}(\sigma_{\rm rel}^{\rm ax})^2}.
\end{eqnarray}
Each $\sigma$ corresponds to the mean atomic wave packet extent.
For instance, we have $\sigma_{\rm cm}^{\rm
rad}=\sqrt{(2\overline{n}_{\rm cm}^{\rm rad}+1)\langle 0 | (r_{\rm
cm}^{\rm rad})^2 |0 \rangle}$, here $\langle 0 | (r_{\rm cm}^{\rm
rad})^2 |0 \rangle$ is the mean extension $\sqrt{\hbar/(2 m
\omega_{\rm cm}^{\rm rad})}$ of the coupled harmonic oscillators
in the ground state, and $\overline{n}$ is the mean phonon number
in a given mode. $m$ is the atomic mass, and $\omega$ the
frequency of the oscillator, which we estimated for all modes
using the spectroscopy on the quadrupolar transition to be
$(\omega_{\rm cm}^{\rm rad},\omega_{\rm rel}^{\rm ax},\omega_{\rm
rel}^{\rm rad})=2\pi(1.5,0.9,1.1)$\,MHz. Taking the mean phonon
number of each mode to be around $12$ for a Doppler cooled
ion-crystal \cite{Slo12}, we get
\begin{eqnarray}
F_{\rm dyn}=0.45.
\end{eqnarray}
In the limit of weak excitation, $\kappa F_{\rm dyn}$ also
directly corresponds to the visibility of the two-ion
interference. The effect of motion-induced decoherence can be reduced by cooling the radial
modes to the motional ground state \cite{Slo12} or by choosing a
forward Raman scattering scenario \cite{Cab99}. The difficulty of
the last option is that light from the Raman excitation can leak
through the detection channel during the excitation.

The effect of the finite coherence times of the individual qubits
is included in the coherence factor $ e^{-4 t/\tau}$.
In our experiment, the coherence of the individual RF qubits is
limited mostly by the ambient magnetic field fluctuations. For
each atom, we measured it to be 120\,$\mu$s. The noise seen by
both ions is correlated when they leave the decoherence free
subspace \cite{Mon11}. For our experiment, this amounts to a
decrease of our coherence on average to about $e^{-4 t/\tau} =
0.96$.

Last, the coherence and overall fidelity of the generated
entangled state is limited by the imperfect populations of the
desired $|ge \rangle$ and $|eg\rangle$ states. This is effectively
accounted for in the overall fidelity by the factor $\kappa$,
which we estimated from the populations measurements to be
0.89$\pm 0.03$. This is in good agreement with the excitation
probability $p_{\rm e}=0.07 \pm 0.03$ of each ion and the measured
dark-counts of our avalanche photodiode of 10\,counts/s.

By inserting all the mentioned inefficiencies into the
Eq.~\ref{fidelity}, we get the overall fidelity of our measured
state with the maximally entangled state $|\Psi^+\rangle$ to be
$F=0.62$, in good agreement with the measured fidelity of $0.64
\pm 0.02$.

\subsection{Measurement of the second-order correlation function}

In order to estimate the degree of the spatial and polarization
indistinguishability of the photons coming from the two ions, we
measure the second order correlation function $g^{(2)}(\tau)$. To
reach a time-resolution beyond the spontaneous decay time, we
implemented a Hanbury Brown and Twiss set-up, by splitting the
fluorescence into two parts with a non-polarizing beam-splitter
and inserting a second avalanche photodiode (APD2). The
unnormalised correlation function $G^{(2)}(\tau)$ reads
\begin{eqnarray}\label{Sec_ord_corr}
G_{\rm Tot}^{(2)}(\tau)=\langle
\hat{E}^-(t)\hat{E}^-(t-\tau)\hat{E}^{+}(t-\tau)\hat{E}^{+}(t)\rangle,
\end{eqnarray}
where $\hat{E}^+$ and $\hat{E}^-$ are positive and negative frequency parts of the field operator $\hat{E}$ and $\tau$ is the time delay between two clicks at the two detectors. Field $\hat{E}=\vec{e}_1\hat{E}_1+\vec{e}_2\hat{E}_2 e^{i\phi}$ corresponds to the coherent sum of the two field operators from atoms $1$ and $2$ with a phase difference $\phi$ and $\vec{e}_1$ and $\vec{e}_2$ are the polarization vectors.
For this measurement, we do not stabilize the optical paths. By
inserting the field expression into equation (\ref{Sec_ord_corr}),
and averaging over the optical phases $\phi$, we get $G_{\rm
Tot}^{(2)}(\tau)=2(G^{(2)}(\tau)+|\vec{e}_1 \vec{e}_2|^2 |G^{(1)}(\tau)|^2+\langle n
\rangle^2)$, where $G^{(1)}(\tau)$ is the (unnormalised) single ion
first order correlation function $G^{(1)}(\tau)=\langle
\hat{E}^-(t)\hat{E}^+(t+\tau)\rangle$ and $\langle n
\rangle$ is the mean number of photons on each APD.

When normalizing the second order correlation to $\langle
\hat{E}^-(t) \hat{E}^{+}(t)\rangle^2$, one obtains
\begin{eqnarray}\label{Sec_ord_corr_cal}
g_{\rm Tot}^{(2)}(\tau)=\frac{1}{2}(g^{(2)}(\tau)+|\vec{e_1}\vec{e_2}|^2|g^{(1)}(\tau)|^2+1),
\end{eqnarray}
where $g^{(1)}$ and $g^{(2)}$ stand for first and second order
normalized correlation functions, respectively. For two single
atoms, $g^{(2)}(0)=0$ (antibunching) and $g^{(1)}(0)=1$, so that total second order correlation function is equal to
$g_{\rm Tot}^{(2)}(0)=\frac{1}{2}(1+|\vec{e_1}\vec{e_2}|^2)$. From here it follows, that for
two indistinguishable polarizations $g_{\rm Tot}^{(2)}(0)=1$. The same analysis applies also for the spatial indistinguishability. In our experiment $g_{\rm Tot}^{(2)}(0)=0.98\pm 0.07$, which shows that indistinguishability conditions are fulfilled and that temporal decoherence in the form
of atomic motion is the main limitation.

\subsection{Efficiency budget}
The efficiency for detecting a single Raman-scattered photon in
our setup was estimated to be $\eta = 8\times\,10^{-4}$. It was
derived from the detection probability of a single Raman scattered
photon given by the collection efficiency of our lenses ($\sim
0.04$), the single-mode fiber coupling efficiency ($\sim 0.1$) and
by the avalanche-photodiode detection efficiency ($\sim 0.6$).
Additional factors of 0.5 and 0.66 come from the polarization
filtering of the unwanted $\pi$-polarized photons and from the
probability of decaying back to the $|g\rangle$ state after the
Raman pulse excitation, respectively. The actual entanglement rate
may thus be estimated. With our single ion excitation probability
$p_e = 0.07\pm 0.03$\%, the overall probability for detecting a
single photon from one of the two ions is then $P_{\rm succ}\sim 2 p_e
\eta = 1.1 \times 10^{-4}$. For comparison, the heralding
entanglement scheme proposed by Simon et al.\cite{Sim03} would
give for our experimental setup approximately $P_{\rm succ}
\approx 2 \eta^2 = 1.3\times 10^{-6}$, so about two orders of magnitude
smaller success probability of entanglement generation in a given
experimental trial. For simplicity, we assumed here $p_e=1$ for the two-photon scheme and an
additional factor of two comes from the two possible contributions
to the coincidence detection events.

\end{document}